\documentclass[preprint]{revtex4}
\usepackage{graphics}
\def \beq{\begin{equation}}
\def \eeq{\end{equation}}
\def \beqa{\begin{eqnarray}}
\def \eeqa{\end{eqnarray}}
\def \ppbar{\langle\overline\psi\psi\rangle}
\def \alphas{\alpha_{\scriptscriptstyle S}}
\def \chil{\chi_{\scriptscriptstyle L}}
\def \chif{\chi_{\scriptscriptstyle F}}
\def \lms{\Lambda_{\overline{MS}}}
\def \avm{{\overline m}}
\def \mpi{m_\pi} \def \mrho{m_\rho}
\def \mpip{m_{\pi^+}} \def \mrhop{m_{\rho^+}}
\def \mpiz{m_{\pi^0}} 
\def \w{\langle {\rm Re}\,L\rangle}
\def \tr{{\rm Tr}\,}
\begin{document}

\title{The phase transition in QCD with broken $SU(2)$ flavour symmetry}
\author{Rajiv V.\ \surname{Gavai}}
\email{gavai@tifr.res.in}
\author{Sourendu \surname{Gupta}}
\email{sgupta@tifr.res.in}
\affiliation{Department of Theoretical Physics, Tata Institute of Fundamental
         Research,\\ Homi Bhabha Road, Mumbai 400005, India.}

\begin{abstract}
We report the first investigation of the QCD transition temperature,
$T_c$, for two flavours of staggered quarks with unequal masses at
lattice spacings of $1/4T$. On changing the $u$/$d$ quark mass ratio
in such a way that $\mpiz^2/m_{\pi^\pm}^2$ changes from 1 to 0.78,
thus bracketing the physical value of this ratio, we find that $T_c$
remains unchanged in units of both $m_\rho$ and $\lms$.
\end{abstract}
\preprint{TIFR/TH/02-27, hep-lat/0208019}
\maketitle

Lattice simulations of QCD with dynamical quarks aim to generate weights
for the Euclidean path integral
\beq
   Z = \int{\cal D}U\exp[-S] {\rm Det}M_u{\rm Det}M_d,
\label{part}\eeq
discretised on a space-time lattice. Here $U$ are gauge fields which
enter the action $S$ and the determinants of the Dirac operator $M$ for
both $u$ and $d$ flavours. This leads to the well-known doubling problem
in the chiral limit. Several solutions are known and have been used fairly
extensively. One is to work with Wilson quarks, for which chiral symmetry
is broken through an irrelevant operator at finite lattice spacing
and recovered in the continuum. Among the new solutions is the domain
wall definition of quarks \cite{dwf} in which the theory is extended to
five dimensions, and by suitable tweaking of this extended action chiral
symmetry is obtained on a four dimensional slice through the lattice when
the length of the fifth dimension is sent to infinity. Another recently
discovered solution is the overlap definition \cite{overlap}, in which
chiral symmetry is intact but its generator and the Dirac operator are
non-local at finite lattice spacing, but become local in the continuum.

In this study we used the solution popular in finite temperature
studies--- staggered quarks. These have an exact continuous chiral
symmetry at all lattice spacings, $a$, and in the continuum limit give
4 degenerate flavours in 4-dimensions.  Two degenerate flavours are
obtained by the prescription
\beq
   {\rm Det}M_u{\rm Det}M_d = \biggl({\rm Det}M_{stag}\biggr)^{1/2},
\label{2fl}\eeq
where $M_{stag}$ is the determinant for a single staggered quark field.

In this paper we introduce two staggered quark fields and define two
flavours by the prescription
\beq
   {\rm Det}M_u{\rm Det}M_d =
        \biggl({\rm Det}M_{stag(u)}{\rm Det}M_{stag(d)}\biggr)^{1/4},
\label{11fl}\eeq
which we call 1+1 flavours.  When the two quark masses are degenerate,
this definition gives the same weight in the partition function of eq.\
(\ref{part}) as the definition in eq.\ (\ref{2fl}).  The new prescription
allows us to handle problems involving breaking of vector $SU(2)$ flavour
symmetry, such as unequal masses for the $u$ and $d$ quarks (which we
explore in this paper) and putting isovector chemical potential on the
lattice \cite{isovector}.

\begin{figure}[hbt]\begin{center}
   \includegraphics{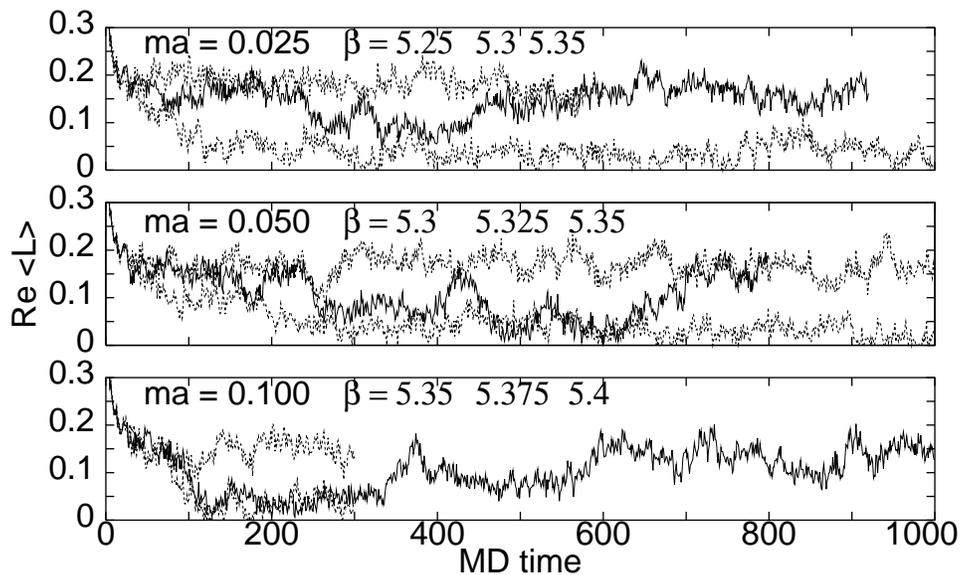}
   \end{center}
   \caption{Run time histories of $\w$ on $4\times 8^3$ lattices for
       1+1 flavours of staggered quarks at several couplings close to
       $\beta_c$. The dark lines correspond to runs at $\beta_c$.
       Of the two lighter lines, the upper always corresponds to a
       simulation at $\beta>\beta_c$. The thermalisation time is not
       more than 100 trajectories. $\beta_c$ can be easily identified
       by eye from the long autocorrelation time $\tau$.}
\label{fg.runtime}\end{figure}

\begin{figure}[ht]\begin{center}
   \scalebox{0.6}{\includegraphics{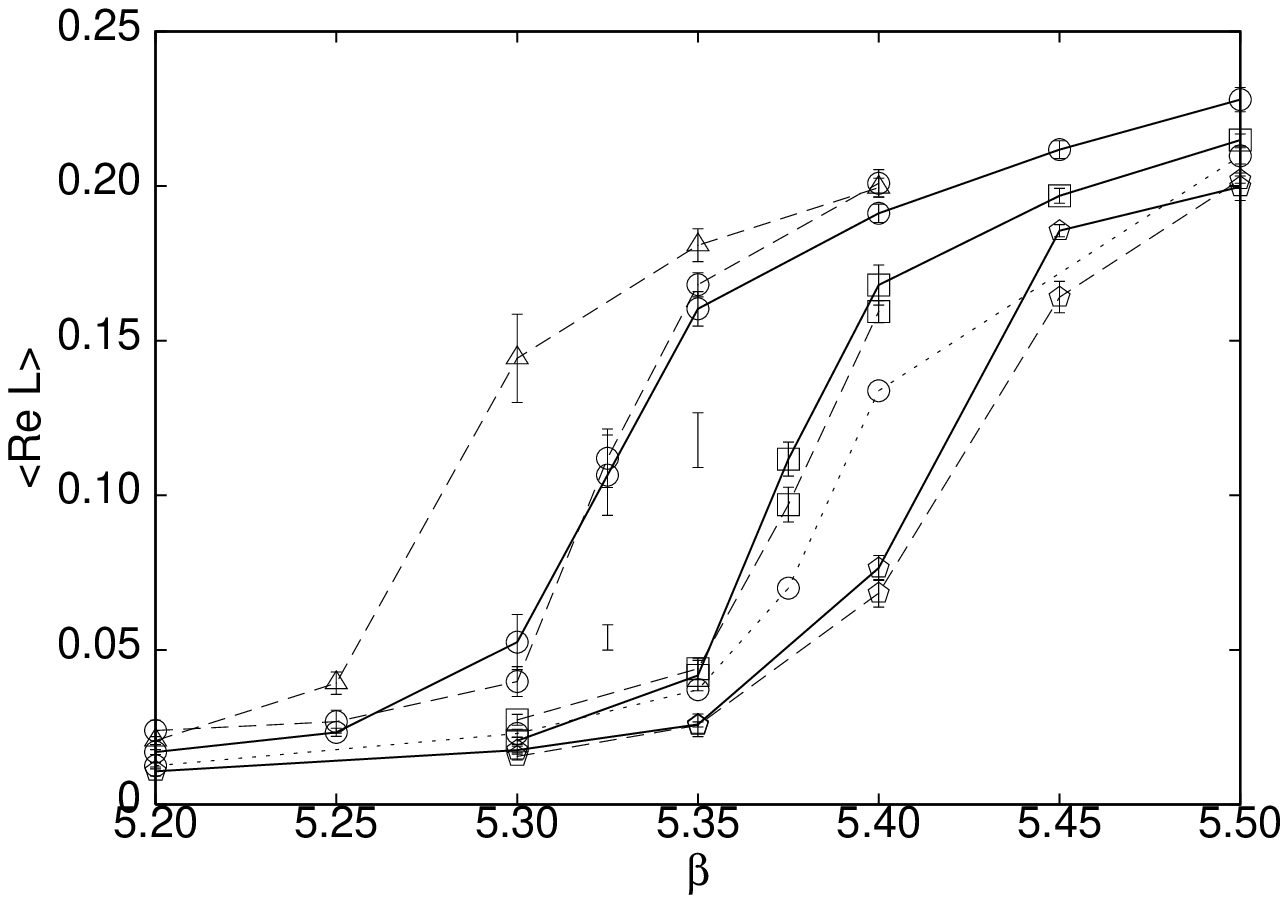}}
   \scalebox{0.6}{\includegraphics{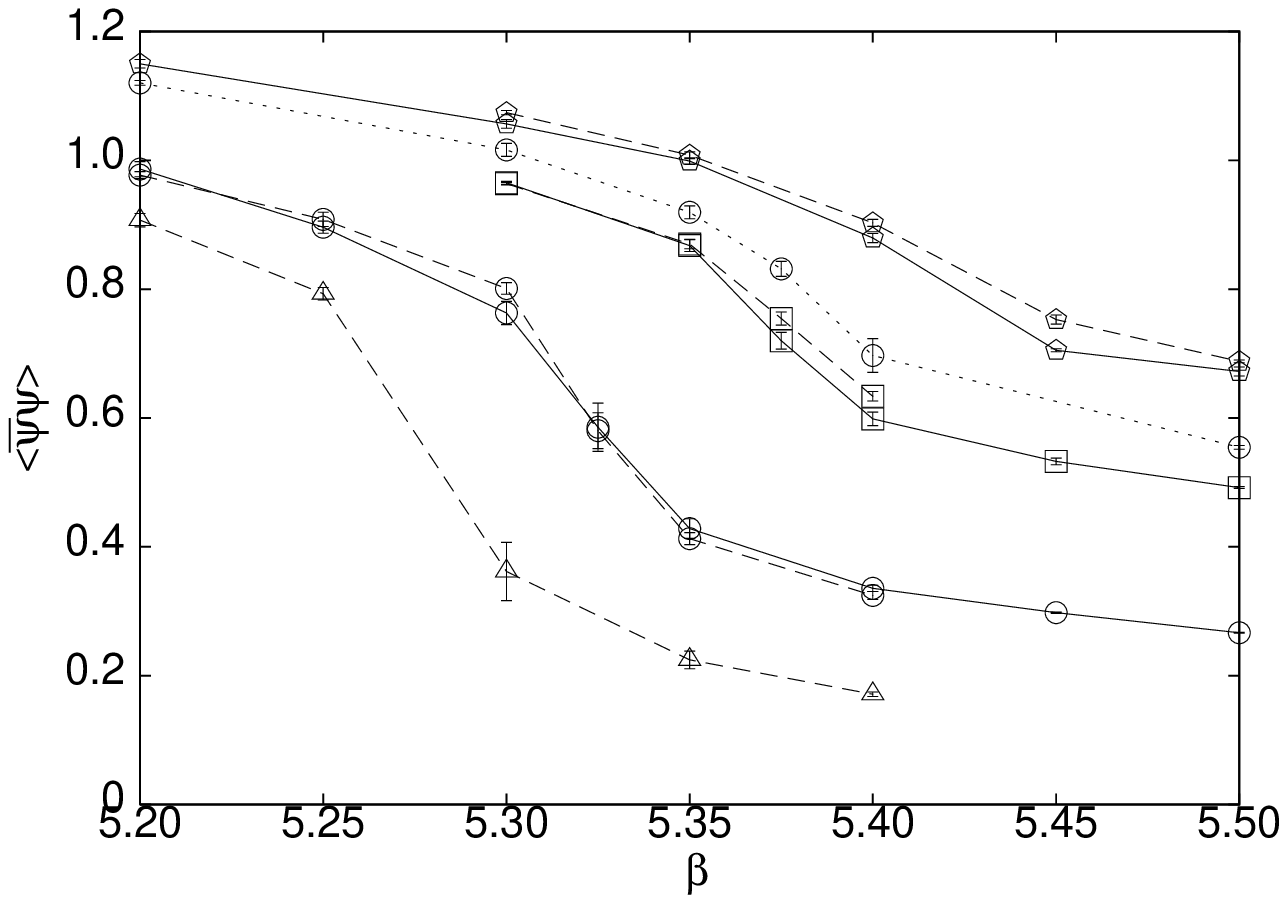}}
   \end{center}
   \caption{$\w$ and $\ppbar$ as functions of $\beta$ for $\avm=0.025$
       (triangles) 0.05 (circles), 0.1 (boxes) and 0.15 (pentagons). The lines
       join centers of the measured data points. There is no significant
       difference between the measurements with $m_d/m_u=1$ (dashed lines) and
       2 (full lines). However, when $m_d/m_u=10$ (dotted line), there is a
       significant shift in $\beta_c$.}
\label{fg.thermo}\end{figure}

\begin{figure}[ht]\begin{center}
   \includegraphics{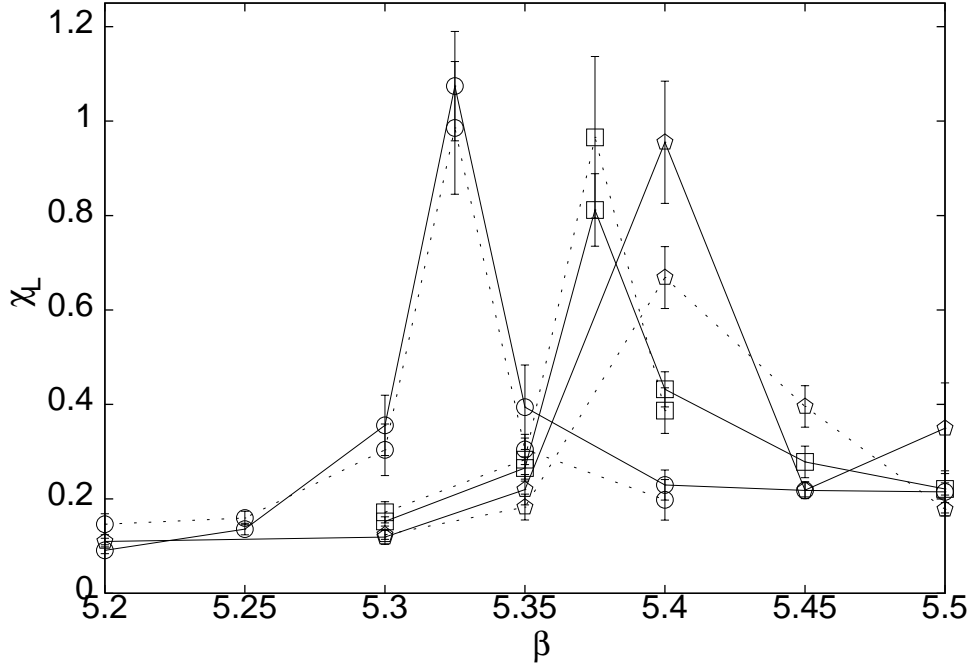}
   \end{center}
   \caption{The Wilson-line susceptibility $\chi_{\scriptscriptstyle L}$ as a
       function of $\beta$ for $\avm=0.05$ (circles), 0.1 (boxes) and 0.15
       (pentagons). The lines join centers of the measured data points for
       $m_d/m_u=2$ (full lines) and 1 (dotted lines).}
\label{fg.chil}\end{figure}

\begin{figure}[ht]\begin{center}
   \includegraphics{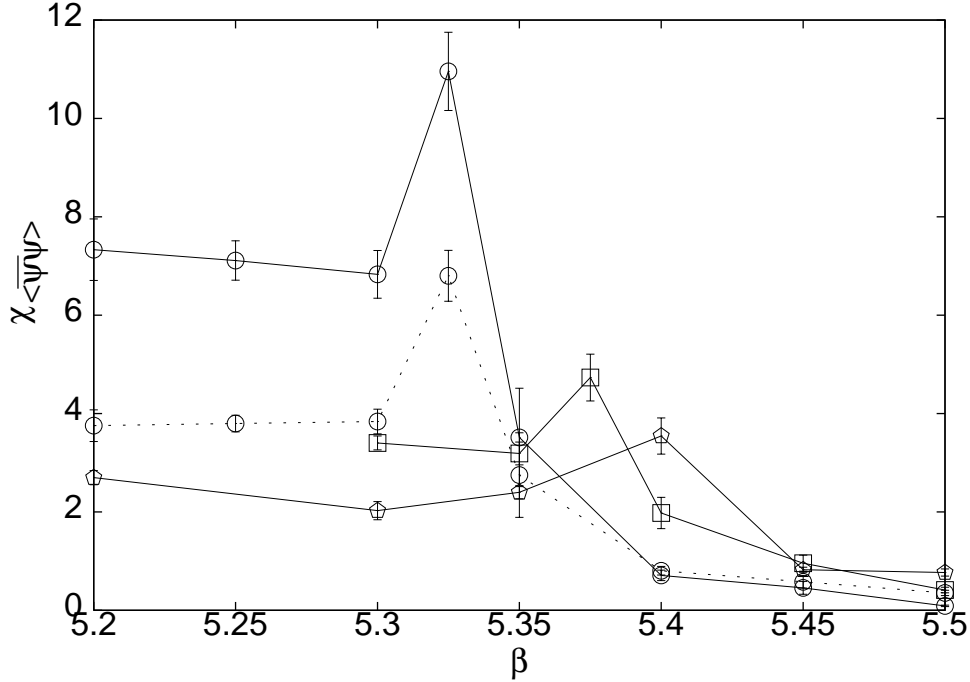}
   \end{center}
   \caption{The susceptibility for the chiral condensate, $\chi_{\ppbar}$,
       as a function of $\beta$ for $\avm=0.05$ (circles), 0.1 (boxes) and
       0.15 (pentagons) and for $m_d/m_u=2$. The lines join centers of the
       measured data points. The $\langle\overline uu\rangle$ susceptibility
       is shown except at the lightest $a\avm$ where the $\langle\overline
       dd\rangle$ susceptibility is also shown for comparison (dotted line).}
\label{fg.chif}\end{figure}

\begin{table}[htbp]
\begin{center}\begin{tabular}{|c|c||l|c|c|c|c|}
 \hline
 $a\avm$ & $m_d/m_u$ & & All & 3:13 & 5:11 & All (2 mass) \\
 \hline
 0.025 & 1 & $am_{\pi}$  & 0.422 (2) & $\underline{0.419 (4)}$ & 0.42 (1) & 0.419  (4) \\
       & 1 & $am_{\rho}$  & $\underline{1.16 (4)}$ & 1.4 (2) & & \\
 \hline
 0.05 & 1 & $am_{\pi}$   & 0.587 (4) & $\underline{0.579 (6)}$ & 0.585 (8) & 0.579  (6) \\
      & 2 & $am_{\pi^+}$ & 0.588 (2) & $0.582 (2)$ & 0.583 (5) & \underline{0.581 (2)} \\
      & 2 & $am_{\pi^0}$ & 0.535 (4) & 0.521 (3) & 0.516 (5) & $\underline{0.516 (4)}$ \\
      & 1 & $am_{\rho}$   & $\underline{1.29 (5)}$ & 1.4 (1) & 1.1 (4) & \\
      & 2 & $am_{\rho^+}$ & $\underline{1.31 (5)}$ & 1.5 (2) & & \\
      & 2 & $am_{\rho^0}$ & $\underline{1.31 (5)}$ & 1.5 (3) & & \\
 \hline
 0.10 & 1 & $am_{\pi}$   & 0.815 (5) & $\underline{0.805 (9)}$ & 0.80  (2) & 0.80  (2) \\
      & 2 & $am_{\pi^+}$ & 0.818 (3) & $\underline{0.811 (2)}$ & 0.811 (3) & 0.811 (2) \\
      & 2 & $am_{\pi^0}$ & 0.740 (6) & 0.717 (3) & 0.707 (4) & $\underline{0.705 (6)}$ \\
      & 1 & $am_{\rho}$   & $\underline{1.41 (5)}$ & & 1.4  (3) & \\
      & 2 & $am_{\rho^+}$ & $\underline{1.39 (4)}$ & & 1.4  (2) & \\
      & 2 & $am_{\rho^0}$ & $\underline{1.38 (4)}$ & & 1.4  (3) & \\
 \hline
 0.15 & 1 & $am_{\pi}$   & 0.973 (2) & 0.972 (3) & 0.978 (4) & \underline{0.970 (3)} \\
      & 2 & $am_{\pi^+}$ & 0.982 (2) & 0.971 (2) & 0.968 (3) & \underline{0.968 (3)} \\
      & 2 & $am_{\pi^0}$ & 0.885 (7) & 0.852 (4) & 0.838 (2) & $\underline{0.823 (7)}$ \\
      & 1 & $am_{\rho}$   & $\underline{1.44 (3)}$ & & 1.32 (3) & \\
      & 2 & $am_{\rho^+}$ & $\underline{1.47 (3)}$ & & 1.47 (5) & \\
      & 2 & $am_{\rho^0}$ & $\underline{1.45 (3)}$ & & 1.50 (3) & \\
 \hline
\end{tabular}\end{center}
\caption{Meson masses from fits to the form shown in eq.\ (\ref{form})
   with one or two masses. The range of $t$ to which the correlation
   function is fitted is indicated in the column header. The entries
   for $m_\pi$ and $m_\rho$ are measurements with degenerate quark
   masses. The remainder are for $m_d=2m_u$. We consider the underlined
   entries to be our best estimates of the meson masses.}
\label{tb.mass}\end{table}

Our strategy here is to proceed in two steps--- first to compare 2
flavours with 1+1 flavours of degenerate light quarks, and next to
lift the degeneracy by giving unequal masses to the quarks (we shall
use the notation $\avm=(m_u+m_d)/2$). In each case we simulate the finite
temperature theory on $4\times8^3$ lattices for several $\avm$. We
study the real part of the Wilson line, $\w$, the quark condensate,
$\ppbar=(\langle\overline uu\rangle + \langle\overline dd\rangle)/2$
and the related susceptibilities
\beqa
\nonumber
   \chil &=& V \left(\left\langle(Re L)^2\right\rangle -
             \left\langle\vphantom{(Re L)^2}Re L\right\rangle^2\right),\\
   \chif &=& V \left(\left\langle(\tr M^{-1})^2\right\rangle -
             \left\langle\vphantom{(\tr M^{-1})^2}\tr M^{-1}\right\rangle^2
               + \left\langle(\tr M^{-2})\right\rangle\right)
          = \chi_m +\chi_\pi,
\label{chil}\eeqa
where $\chi_m$ is the chiral susceptibility defined in \cite{bi} and $\chi_\pi$
is the pion susceptibility of \cite{sg}. $\chif$ has the simple
stochastic estimator \cite{pushan}
\beq
   \chif/V = \left\langle\overline{(r^\dag M^{-1}r)^2}\right\rangle -
      \left\langle\vphantom{(r^\dag M^{-1}r)^2}\overline{r^\dag M^{-1}r}\right\rangle^2,
\label{stoch}\eeq
where each $r$ is a complex random vector with each component drawn from
a Gaussian distribution of unit width and the bar denotes an average over
such an ensemble.  We also study the integrated autocorrelation time,
$\tau$ \cite{auto}, of $\w$ and $\ppbar$, measured self-consistently over
runs of length of at least $30\tau$. $\beta_c$ can be defined by peaks in
$\chil$, $\chif$ and $\tau$. Within the accuracy of our determinations,
they agree.  We also make zero temperature measurements of the plaquette,
$P$, and the masses of the pion, $\mpi$, and the rho meson, $\mrho$, at
$\beta_c$ on $8^4$ lattices (we confirm that the lattice is large enough
for these measurements by checking that the zero temperature measurements
are independent of boundary conditions).  The measurement of $P$ at $T=0$
also allows us to extract $\alphas$ \cite{blm} and hence to convert $a$
into a physical length scale in order to extract $T_c/\lms$.

Our measurements have been taken at bare quark masses, $a\avm=0.15$,
$0.1$, $0.05$ and $0.025$ for degenerate flavours. For
non-degenerate flavours with $m_d=2m_u$ we have only worked with the
first three values of $\avm$. We have also investigated $m_d=10m_u$
with $a\avm=0.15$. The simulations have been performed with the
Hybrid-R algorithm \cite{ralg}. This algorithm is based on a molecular
dynamics (MD) evolution which treats degenerate 1+1 and 2 flavours
differently. Hence, the agreement in the results of these two cases,
which we show later, is a good test that the numerical treatment of the
discretised MD evolution is free of errors at the level of accuracy
we achieve. The trajectory lengths have been taken to be one unit of
molecular dynamics time, and the MD equations have been integrated over
this range in 100 steps. Some typical run time histories of $\w$ are
shown in Figure \ref{fg.runtime}. Run time histories of other quantities
are similar. The slow but large fluctuations visible in some of these
runs are typical of critical slowing down, and therefore yield estimates
of $\beta_c$.

Our measurements of $\w$ and $\ppbar$ are collected in Figure \ref{fg.thermo}.
Clearly there are no statistically significant differences between
the measurements with degenerate quark masses and for $m_d/m_u=2$.
From these figures it is also clear that $\beta_c$ is not sensitive
to variation of $m_d/m_u$ in the range 1--2. From our measurements of
masses, detailed later, it turns out this range of bare quark mass ratios
includes the range of $\mpiz^2/\mpip^2$ between 1 and roughly 0.8. Since
the physical pion masses \cite{pdg} yield $\mpiz^2/\mpip^2=0.935$,
which is inside this range, we may conclude that the shift in $\beta_c$
due to realistic vector $SU(2)$ flavour symmetry breaking is negligible
at the level of accuracy we can reach.  A significant downward shift in
$\beta_c$ is seen when the ratio $m_d/m_u$ is made as large as 10,

The measurements of autocorrelations yield strong peaks, sometimes as high
as 300 MD time units, at couplings which we identify as $\beta_c$. These
runs were extremely time consuming, since some of them required more
than $10^4$ trajectories for reliable estimates of the susceptibilities
shown in Figures \ref{fg.chil} and \ref{fg.chif}. These also peaked at the
same values of $\beta$, and are our primary means for the identification
of $\beta_c$. Very good agreement between measurements of $\chil$ for
$m_d/m_u=2$ and 1 indicate that there is no shift in $\beta_c$.  The only
exception is the value of $\chil$ near $\beta_c$ for $a\avm=0.15$,
where the difference can be attributed to the factor two difference in
statistics between the runs with $m_d/m_u=2$ and 1--- the former have both
larger statistics and $\chil$. In Figure \ref{fg.chif} we show $\chif$,
which also peak at the same $\beta_c$, but unlike $\chil$ have stronger
quark mass dependence.

Our estimates of $\beta_c$ for 1+1 flavours are shown in Table
\ref{tb.summary}.  Measurements of $\beta_c$ with the 2 flavour definition
of staggered quarks have been made earlier in \cite{gott} on $4\times8^3$
lattices with $a\avm=0.1$, $0.05$ and $0.025$. Estimates of $\beta_c$
have also been made on lattices with $N_t=4$ for $a\avm=0.025$ and
$0.0125$ in \cite{fuku}. These earlier results are fully in agreement
with our measurements.  Recall that studies on fixed volumes cannot
decide the order of the transition, or indeed whether the peaking of the
susceptibilities are due to a phase transition or a cross-over. That
requires a study with varying volumes--- a work that is outside the
scope of this investigation and left for later.

We turn now to measurements at zero temperature. For this part of the
work we generated configurations on $8^4$ lattices at the coupling
$\beta_c$ for the value of $a\avm$ under study. The autocorrelations
of $\ppbar$ were found to be less than 5 MD time units in all cases,
and the thermalisation time was less than 50 trajectories. We discarded
the first 50 trajectories and stored 50 configurations separated by 5
trajectories in each of these simulations.

Meson correlators were computed on the stored configurations by inverting
the staggered Dirac operator with appropriate masses. For $m_u\ne m_d$,
the flavour combinations which give the meson propagators are---
\beqa
\nonumber
   C_{\pi^+}(r) &=& \langle M_u^{-1}(0,r) M_d^{-1}(r,0) \rangle \\
   C_{\pi^0}(r) &=& \frac12 \biggl[
                    \langle M_u^{-1}(0,r) M_u^{-1}(r,0) \rangle
                +   \langle M_d^{-1}(0,r) M_d^{-1}(r,0) \rangle \biggr],
\label{pion}\eeqa
where $r$ labels a point of the lattice with respect to the chosen origin.
The correlator for $\pi^-$ is obtained by flipping the roles of the $u$
and $d$ quarks in the expectation value on the right.  The analogous
combinations for the $\rho$ have the usual staggered fermion phase
factor. The correlators, and hence the masses, for opposite charge states
are identical.  For degenerate quark masses, the propagators in all the
charge states are identical.  As usual, these correlators are summed
over spatial slices to give the zero momentum propagators.

\begin{table}[htb]
\begin{center}\begin{tabular}{|c|c|c|c|c|c|}
 \hline
 $a\avm$ & $\beta_c$ & $\mpip^2/\mrhop^2$ & $\mpiz^2/\mpip^2$ & $T_c/\mrhop$ &
           $T_c/\lms$ \\
 \hline
 0.050 & 5.325 (25) & 0.20 (2)  & 0.78 (2) & 0.194 (7) & 1.2 (2) \\
 0.100 & 5.375 (25) & 0.34 (2)  & 0.76 (1) & 0.177 (6) & 1.3 (2) \\
 0.150 & 5.400 (25) & 0.44 (2)  & 0.72 (1) & 0.172 (4) & 1.3 (2) \\
 \hline
\end{tabular}\end{center}
\caption{A summary of our measurements with $m_d=2m_u$. These measurements
   are statistically indistinguishable from measurements with $m_d=m_u$
   (except for the ratio $\mpiz^2/\mpip^2$ which is then identically unity).
   In the limit of physical $\avm$, $T_c=175\pm6$ from extrapolation of
   $T_c/\mrho$ and $T_c=167\pm9^{+15}_{-14}$ MeV from extrapolation of
   $T_c/\lms$.}
\label{tb.summary}\end{table}

Meson masses were obtained by fitting these zero momentum
correlators to the form
\beq
   C(t) = A\cosh\left[m\left(\frac L2-t\right)\right]
        + A'\cosh\left[m'\left(\frac L2-t\right)\right],
\label{form}\eeq
or the corresponding single mass formula obtained by dropping the second
term. We explored the stability of the fits by changing the range of
$t$ over which the single mass form was fitted, and by comparing the
result with the two mass fits.

The detailed comparisons are given in Table \ref{tb.mass}.  For the
flavour symmetric pion and the $\pi^+$, there is a clear stable mass for
fits over the range $3\le at\le13$. For the $\pi^0$ the mass extracted
from the two mass fit lies below that obtained from single mass fits. In
all cases we took the lowest estimate to be the value of the mass. For
the various $\rho$, the fit errors are larger and we accept the single
mass fit over the full range of $t$ as our best estimate of the mass. The
1+1 flavour masses obtained with $m_u=m_d$ are fully consistent with
previous measurements for 2 flavours \cite{gott}.

We draw attention to the fact that the mass splitting between the
charged and uncharged $\rho$ is not visible within measurement errors,
and both these masses are equal to that obtained with degenerate quarks.
However, the splitting between the neutral and charged pions is clearly
visible. Interestingly, the latter are statistically indistinguishable
from the pion mass measured with equal quark masses. Breaking $SU(2)$
flavour symmetry only results in lowering the $\pi^0$ mass.

From our extraction of $a\mrho$, we extracted the ratio
$T_c/\mrho=1/N_ta\mrho$. These estimates are collected in Table
\ref{tb.summary}.  The ratio $T_c/\lms$ increases marginally as $a\avm$
decreases, due to the small decrease in the measured values of $a\mrho$,
We have extrapolated these estimates to the values of $a\avm$ for which
the physical value of $\mpi/\mrho$ is obtained, through the form---
\beq
   m_\rho = a + b \avm + \cdots.
\label{chiral}\eeq
For the value of $a\avm$ where the physical value of $\mpi/\mrho$ is
obtained, we find $T_c/\mrho=0.227\pm0.008$. This gives $T_c=175\pm6$
MeV. Mutually consistent values of $T_c$ are obtained using the 2-flavour
data of \cite{gott}, and our 1+1 flavour data with $m_d/m_u=1$ as well
as 2.

\begin{figure}[hbt]\begin{center}
   \includegraphics{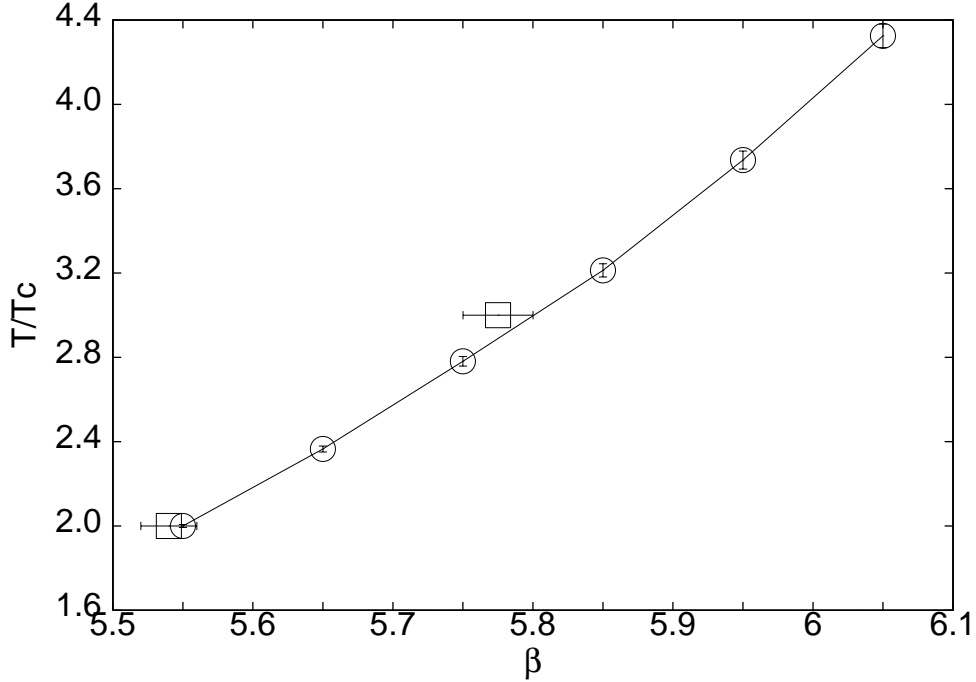}
   \end{center}
   \caption{The relation between the lattice spacing and $\beta$, shown
       in terms of the dependence of $T/T_c$ on $\beta$ for $N_t=4$ lattices.
       The circles denote estimates from measurements of $P$. Squares denote
       estimates from measurements of $\beta_c$ on lattices with different
       $N_t$ \cite{milc}.}
\label{fg.scale}\end{figure}

Alternatively, the lattice spacing can be traded for $\lms$ by using
our measurements of the plaquette, $P$, to extract the running coupling,
$\alphas$, and using this in the 2-loop $\beta$ function. This procedure
was introduced in \cite{blm} and later used to determine $T_c/\lms$ from
the 2 flavour data for realistic quark masses in \cite{tclam}. With such
an analysis, the values of $T_c/\lms$ extracted from our measurements in
the 1+1 flavour case agree completely with the previous values at the same
$a\avm$.  Extrapolation of our 1+1 flavour results to quark masses which
give the correct value of $\mrho/\lms$, yield $T_c/\lms=0.49\pm0.02$, as
in \cite{tclam}. Then, using the 2-loop value of $\lms=343^{+31}_{-28}$,
as appropriate below the charm quark threshold \cite{pdg}, we obtain
$T_c=167\pm9^{+15}_{-14}$ MeV (where the first error comes from $T_c/\lms$
and the second from $\lms$). The two estimates of $T_c$ are totally
compatible with each other, and also from other estimates using improved
actions \cite{improved}.

The above method depends on the expansion of $P$ at $T=0$ in a series in
$\alphas$.  Since this is scheme dependent \cite{schemes}, the extraction
of $T_c/\lms$ can have large uncertainties if the cutoff is large. To
estimate these, we have extrapolated the temperature scale
to smaller lattice spacings and compared to direct measurements of the
scale.  At the smallest $\avm=0.025/a$ we have made a few $T=0$ runs at
larger $\beta$ to estimate the lattice spacing through a measurement of
$P$. Since the lattice spacing, $a$, is an outcome of this computation,
a series of runs is needed to tune $\avm$. The scale determined in
this way can be converted into a temperature scale for simulations
with $N_t=4$ at the corresponding $\beta$. This is shown in Figure
\ref{fg.scale}. The temperature scale for $N_t=4$ can also be calibrated
by direct measurements of $\beta_c$ at larger $N_t$. For 2 flavours
of quarks and $\avm=0.025/a$, such measurements have been performed
\cite{milc}. These results are also plotted in Figure \ref{fg.scale}. The
good agreement between the two methods of setting the scale implies that
the lattice spacings are fine enough for 2-loop scaling to work. As a
result, we expect that $T_c/\lms$ obtained here are relevant
to the continuum limit.

The major remaining uncertainties are in the extrapolation to zero quark
mass, and in possible power corrections in $a$ to various quantities
we have measured.  The question of the order of the phase transition
needs a finite size scaling study and has not been addressed here.
A detailed study of these issues lies outside the scope of this paper,
and is left to the future.


\begin{thebibliography}{99}
\bibitem{dwf}
   D.\ Kaplan, {\sl Phys.\ Lett.\/}, B 288 (1992) 342;\\
   Y.\ Shamir, {\sl Nucl.\ Phys.\/}, B 406 (1993) 90.
\bibitem{overlap}
   H.\ Neuberger and R.\ Narayanan, {\sl Phys.\ Rev.\ Lett.\/}, 71 (1993) 3251.
\bibitem{isovector}
   J.\ B.\ Kogut and D.\ K.\ Sinclair, hep-lat/0201017;
   S.\ Gupta, hep-lat/0202005.
\bibitem{bi}
   S.\ Aoki {\sl et al.\/}, {\sl Phys.\ Rev.\/}, D 57 (1998) 3910;\\
   E.\ Laermann, {\sl Nucl.\ Phys.\ Proc.\ Suppl.\/}, 63 (1998) 114.
\bibitem{sg}
   S.\  Gupta, {\sl Phys.\ Lett.\/}, B 286 (1992) 112.
\bibitem{pushan}
   R.\ V.\ Gavai {\sl et al.\/}, {\sl Phys.\ Rev.\/}, D 65 (2002) 054506.
\bibitem{auto}
   For the definition of the integrated autocorrelation time, see, for example,
   A.\ Billoire {\sl et al.\/}, {\sl Nucl.\ Phys.\/}, B 358 (1991) 231.
\bibitem{blm}
   G.\ P.\ Lepage and P.\ B.\ Mackenzie, {\sl Phys.\ Rev.\/} D 48 (1993) 2250.
\bibitem{ralg}
   S.\ Gottlieb {\sl et al.\/}, {\sl Phys.\ Rev.\/} D 35 (1987) 2531.
\bibitem{gott}
   S.\ Gottlieb {\sl et al.\/}, {\sl Phys.\ Rev.\ Lett.\/}, 59 (1987) 1513.
\bibitem{fuku}
   M.\ Fukugita {\sl et al.\/}, {\sl Phys.\ Rev.\ Lett.\/}, 65 (1990) 816;\\
   M.\ Fukugita {\sl et al.\/}, {\sl Phys.\ Rev.\/}, D 42 (1990) 2936.
\bibitem{tclam}
   S.\ Gupta, {\sl Phys.\ Rev.\/}, D 64 (2001) 034507.
\bibitem{schemes}
   G.\ Martinelli {\sl et al.\/}, {\sl Phys.\ Lett.\/}, 100 B (1981) 485;
   S.\ J.\ Brodsky, G.\ P.\ Lepage and P.\ B.\ Mackenzie,
      {\sl Phys.\ Rev.\/} D 28 (1983) 228.
\bibitem{pdg}
   Particle Data Group, {\sl Eur.\ Phys.\ J.\/}, C 3 (1998) 1.
\bibitem{improved}
   F.\ Karsch {\sl et al.\/}, {\sl Nucl.\ Phys.\/}, B 605 (2001) 579;
   A.\ Ali Khan {\sl et al.\/}, {\sl Phys.\ Rev.\/}, D 63 (2001) 034502.
\bibitem{milc}
   S.\ Gottlieb {\sl et al.\/}, {\sl Phys.\ Rev.\/}, D 47 (1993) 3619;
   C.\ Bernard {\sl et al.\/}, {\sl Phys.\ Rev.\/}, D 54 (1996) 4585.
\end{thebibliography}
\end{document}